\documentclass[12pt,a4paper]{article}
\usepackage{graphics}
\usepackage{amssymb}
\usepackage[T1]{fontenc}
\usepackage[latin1]{inputenc}
\begin{document}
\begin{center} {\bf On a Generalized Fifth-Order Integrable Evolution Equation and its Hierarchy} \\
\vskip 0.5 cm
{\small Amitava Choudhuri$^a$, B. Talukdar$^a$ and S. B. Datta$^b$}\\
{\small $^a$\it Department of Physics, Visva-Bharati University, Santiniketan 731235, India}
{\small $^b$\it Abhedananda Mahavidyalaya, Sainthia 731234, India}\\
\vskip 0.2 cm
{\small \it e-mail : binoy123@sancharnet.in}
\end{center}
\vskip 1.0 cm
A general form of the fifth-order nonlinear evolution equation is 
considered. Helmholtz solution of the inverse variational problem is 
used to derive conditions under which  this equation admits an 
analytic representation. A Lennard 
type recursion operator  is then employed to construct a hierarchy 
of Lagrangian equations. It is explicitly demonstrated that the 
constructed system of equations has a Lax representation and two 
compatible Hamiltonian structures. The homogeneous balance method is used to
derive analytic soliton solutions of the third- and fifth-order equations. 
\newpage
\noindent{\bf\large I.  Introduction}
\vskip 0.3 cm
In recent years studies in fifth-order nonlinear evolution equations 
have received considerable attention primarily because these 
equations possess a host of connections with other integrable equations which
play a role in diverse areas of physics ranging from nonlinear optics {[}1{]}
to Bose-Einstein condensation {[}2{]}. For example, \"Ozer and D\"oken {[}3{]}
used a multiple-scale method  to derive the fifth-order Korteweg-de Vries
(KdV) equation from the higher-order nonlinear Sch\"odinger equation. On the
other hand, a similar method could also be used {[}4{]} to obtain the
nonlinear Sch\"odinger equation from fifth-order KdV flow {[}5{]},
Sawada-Kotera equation {[}6{]} and Kaup-Kupershmidt equation {[}7{]}.\par
Third-order evolution equations can often be solved either by the use of an
inverse spectral method or by taking recourse to a simple change of variables.
This is true for both the linearly dispersive KdV equation and the nonlinearly
dispersive Rosenau-Hymann equation {[}8{]}. In contrast, it is quite difficult
to obtain solutions of the fifth-order equations. This might be another point of interest for recent studies {[}9{]} in these equations.\\\par In this work we derive the condition under which the general fifth-order nonlinear evolution equation
\begin{equation}
u_{t}=u_{5x}+Auu_{3x}+Bu_{x}u_{2x}+Cu^2u_{x}  ,\,\,\,u=x(x,t)   
\end{equation} 
admits an analytic representation {[}10{]} or follows from a Lagrangian. Here
$A$, $B$ and $C$ are constant model parameters. The subscripts on $u$ denote
partial derivatives with respect to that variable and, in particular,
$u_{nx}=\frac{\partial^n u}{\partial x^n}$. We use the fifth-order Lagrangian
equation to define an integrable hierarchy. Further, we provide a Lax
representation {[}5{]} and construct a bi-Hamiltonian structure {[}11{]} for
the system. The Lagrangian approach to nonlinear evolution equation has two
novel features. First, from the Lagrangians or Lagrangian densities we can
construct Hamiltonian densities {[}12{]} which form the set of involutive
conserved densities of the system. Second, the expression for the Lagrangian
represents a useful basis to construct an approximate solution for the
evolution equation {[}13,14{]}. We shall, however, use a direct method {[}15{]}
to obtain explicit analytic soliton solutions. \par In Sec.II we deal with the
inverse variational problem for $(1)$ and derive  relations between the model
parameters for the equation to be Lagrangian. We then make use of an appropriate
pseudo-differential operator to construct a hierarchy of equations and
present results for the first few members of the hierarchy. In Sec.III we
find their Lax representations and examine the bi-Hamiltonian structure. The
results presented are expected to serve as a useful test of integrability. We
devote Sec.IV to present explicit solitonic solutions by using the
homogeneous balance method (HB). We present some concluding remarks in
Sec.V.
\vskip 0.2 cm
\noindent{\bf\large II. Lagrangian system of equations}
\vskip 0.3 cm 
\par In the calculus of variation one is
concerned with two types of problems, namely, the direct and the inverse
problem of Newtonian mechanics. The direct problem is essentially the
conventional one in which one first assigns a Lagrangian and then computes the
equations of motion through Lagrange's equations. As opposed to this, the
inverse problems begins with the equation of motion and then constructs a
Lagrangian consistent with the variational principle {[}10{]}. The inverse
problem of the calculus of variation was solved by Helmholtz {[}16{]} during
the end of the nineteenth century. For continuum mechanics the Helmholtz
version of the inverse problem proceeds by considering an r-tuple of
differentiable functions written as
\begin{equation}
P[v]=P(x,v^{(n)})\,\, \epsilon \,\,   {\cal A}^r
\end{equation}
and then defining the so-called Fr\'echet derivative. The Fr\'echet derivative
of P is the differential operator $D_P:\,\, {\cal A}^q\rightarrow {\cal A}^r$ and is given by 
\begin{equation}
D_P(Q)={\frac{d}{d\epsilon}}\vline_{_{_{_{_{\epsilon=0}}}}}P[v+\,\epsilon\, Q[v]]
\end{equation}
for any $Q \,\,\epsilon \,\,{\cal A}^q$. The Helmholtz condition asserts that P is the Euler-Lagrange expression for some variational problem iff $D_P$ is self-adjoint. When self-adjointness is guaranteed, a Lagrangian density for P can be explicitly constructed using the homotopy formula
\begin{equation}
{\cal L}[v]=\int_{0}^{1}vP[\lambda v]d\lambda\,\,.
\end{equation}
In the following we shall demand the Helmholtz condition to be valid for (1).
This will provide us with certain constraints between the model parameters of
(1) to follow from a Lagrangian density.\par A single evolution equation
$u_t=P[u]$, $  u\,\,\epsilon\,\,\mathbb{R}  $ is never the Euler-Lagrange
equation of a variational problem {[}16{]}. One common trick to put a single
evolution equation into a variational form is to replace $u$ by a potential function
\begin{equation}
u=-w_x,\,\, w=w(x,t)\,\,.
\end{equation}
The function $w$ is often called the Casimir potential. In terms of the Casimir potential, (1) reads
\begin{equation}
w_{xt}=P[w_x]\,\,,
\end{equation}
where 
\begin{equation}
P[w_x]=w_{6x}-Aw_xw_{4x}-Bw_{2x}w_{3x}+C{w_x}^2w_{2x}\,\,.
\end{equation}
From (3) and (7) we obtain
\begin{eqnarray}
D_P=D_{6x}-Aw_xD_{4x}-Aw_{4x}D_x-Bw_{2x}D_{3x}-Bw_{3x}D_{2x}\nonumber \\+C{w_x}^2D_{2x}+2Cw_xw_{2x}D_x \,\,.
\end{eqnarray}
To construct the adjoint operator $ D_{p}^{*}$ of the above Fr\'echet derivative we rewrite (8) as
\begin{equation}
D_P=\sum_{j}P_j[w_x]D_j
\end{equation}
and make use of the definition {[}16{]}
\begin{equation}
D^{*}_{P}=\sum(-D)_j.P_j
\end{equation}
meaning that for any $Q\,\,\epsilon\,\, \cal A$
\begin{equation}
D^{*}_{P}Q=\sum_{j}(-D)_j[P_jQ]\,\,.
\end{equation}
This gives
\begin{eqnarray}
D^{*}_{P}=D_{6x}-Aw_xD_{4x}-(3A-B)w_{4x}D_x-(4A-B)w_{2x}D_{3x}\nonumber \\-(6A-2B)w_{3x}D_{2x}+Cw_{x}^{2}D_{2x}+2Cw_xw_{2x}D_x\,\,.
\end{eqnarray}
Demanding variational self-adjointness we obtain from (8) and (12)
\begin{equation}
B=2A
\end{equation}
while $C$ remains unrestricted. Thus the nonlinear equation 
\begin{equation}
u_{t}=u_{5x}+Auu_{3x}+2Au_{x}u_{2x}+Cu^{2}u_{x} 
\end{equation}
forms a Lagrangian system. We note that the Lax equation {[}5{]} with $A=10$,
$B=20$ and $C=30$ and the Ito equation {[}17{]} with $A=3$, $B=6$ and $C=2$
are of the form (14) while the Sawada-Kotera equation with $A=B=C=5$ and the
Kaup-Kupershmidt equation with $A=10$, $B=25$ and $C=20$ are
non-Lagrangian.\par We now use the fifth-order Lagrangian equation (14) to
define an integrable hierarchy. To that end we introduce a pseudo-differential
or integro-differential operator $\Lambda$ which acts on a generic function
$f(x)$ to give {[}18{]}
\begin{equation}
\Lambda f(x)=f_{xx}-puf(x)+qu_x\int_{x}^{+\infty} dyf(y).
\end{equation}
Further, we introduce a function $g_{x}^{(n)}$ to follow from
\begin{equation}
{\Lambda}^nu_x(x,t)=g_{x}^{(n)},n=0,1,2....\,\,\,\,.
\end{equation}
Here $g_{x}^{(n)}$ is a polynomial in $u$ and its $x$-derivatives (up to derivative of order 2n). Using $f(x)=u_x(x,t)$ in (15) we have 
\begin{equation}
\Lambda f(x)=(u_{2x}-\frac{p+q}{2}u^2)_x\,\,.
\end{equation}
From (16) and (17)
\begin{equation}
{\Lambda}^2u_x(x,t)=u_{5x}-(2p+q)uu_{3x}-(3p+4q)u_xu_{2x}+(p+q)(p+\frac{q}{2})u^2u_x\,\,.
\end{equation}
Comparing (14) and (18) and identifying ${\Lambda}^2u_x(x,t)$ as $u_t$ we can express $p$ and $q$ in terms of $A$. This allows us to write 
\begin{equation}
C=(p+q)(p+\frac{q}{2})=\frac{3A^2}{10}\,\,.
\end{equation}
Therefore, the general form of the fifth-order Lagrangian equation generated by $\Lambda$ via (16) has the form
\begin{equation}
u_{t}=u_{5x}+Auu_{3x}+2Au_{x}u_{2x}+\frac{3A^2}{10}u^2u_{x}\,\,.
\end{equation}
\par We have used (16) to generate a hierarchy of nonlinear evolution equations for $n=0,1,2,3$ etc. The first member of the hierarchy ($n=0$) is a linear equation given by
\begin{equation}
u_t=u_x\,\,,
\end{equation}
while the second one ($n=1$) is a third-order nonlinear equation
\begin{equation}
u_t=u_{3x}+\frac{3A}{5}uu_x\,\,.
\end{equation}
The third member ($n=2$) is obviously the fifth-order equation given in (20).
The corresponding seventh and ninth order equations are given by
\begin{eqnarray}
u_t=u_{7x}+\frac{7A}{5}u_{5x}u+\frac{21A}{5}u_{4x}u_x+7Au_{3x}u_{2x}+\frac{7A^2}{10}u^2u_{3x}\nonumber\\+\frac{14A^2}{5}uu_xu_{2x}+\frac{7A^2}{10}{u_x}^3+\frac{7A^3}{50}u^3u_x 
\end{eqnarray}
and
\begin{eqnarray}
u_t=u_{9x}+\frac{9A}{5}u_{7x}u+\frac{36A}{5}u_{6x}u_x+\frac{84A}{5}u_{5x}u_{2x}+\frac{126A}{5}u_{4x}u_{3x}+\frac{651A^2}{50}u_xu_{2x}^{2}\nonumber\\+\frac{483A^2}{50}u_{x}^{2}u_{3x}+\frac{63A^2}{5}uu_{2x}u_{3x}+\frac{378A^2}{50}uu_xu_{4x}+\frac{63A^2}{50}u^{2}u_{5x}\nonumber\\+\frac{63A^3}{50}uu_{x}^{3}+\frac{126A^3}{50}u^2u_xu_{2x}+\frac{21A^3}{50}u^3u_{3x}+\frac{63A^4}{1000}u^4u_x.
\end{eqnarray}
\skip 0.2 cm
{\bf\large III. Lax representation and bi-Hamiltonian structure}\vskip 0.3 cm\par Integrable nonlinear evolution equations admit zero curvature or Lax
representation {[}5{]}. These equations are characterized by an infinite
number of conserved densities which are in involution. Moreover, each number
of the hierarchy has a bi-Hamiltonian structure {[}11{]}.In the following we
demonstrate these three important features for our equations in (20)-(24).\par
The Lax representation of nonlinear evolution equations is based on the
algebra of differential operators. Here one considers two linear operators $L$
and $M$ . The eigenvalue equation for the operator $L$ is given by
\begin{equation}
L\psi=\lambda\psi
\end{equation}
with $\psi$, the eigenfunction and $\lambda$, the corresponding eigenvalue. The operator $M$ characterizes the change of eigenfunctions with the parameter $t$ which, in a nonlinear evolution equation, usually corresponds to the time . The general form of this equation is 
\begin{equation}
{\psi}_t=M\psi\,\,.
\end{equation}
If we now invoke a basic result of the inverse spectral method that
$\frac{d\lambda}{dt}=0$ for non-zero eigenfunctions {[}19{]}, then (25) and (26) will immediately give
\begin{equation}
\frac{\partial L}{\partial t}=[M,L]\,\,.
\end{equation}
Equation (27) is called the Lax equation and $L$ and $M$ are called the Lax
pairs. In the context of Lax's method it is often said that $L$ defines the
original spectral problem while $M$ represents an auxiliary spectral problem.
For a given nonlinear evolution equation one needs to find these operators.
This is not always a straightforward task. In fact, no systematic procedure
has been derived to determine whether a nonlinear partial differential
equation can be represented in the form (27).\par We shall now find the Lax
representation for the hierarchy of equations given in (20)-(24). We first
note that  as one goes along the hierarchy the original spectral problem
remains invariant while the auxiliary  spectral problem goes on changing.
Keeping this in mind we take 
\begin{equation}
\!\!\!\!\!\!\!\!\!\!\!\!\!\!\!\!\!\!\!\!\!\!\!\!\!\!\!\!\!\!\!\!\!\!\!\!\!\!\!\!\!\!\!\!\!\!\!\!\!\!\!\!\!\!\!\! L=\partial_{x}^{2}+\frac{A}{10}u
\end{equation}
In writing (28) we have exploited the similarity between (22) and the KdV
equation. As regards the auxiliary spectral problem we postulate that for an
evolution equation of the form $u_t=K[u]$ the terms in the Fr\'echet
derivative  of $K[u]$  contribute additively with unequal weights to form the
operator $M$ such that $L$ and $M$ via (22) reproduces $K[u]$ . Of course,
there should not be any inconsistency in determining the values of the weight factors. For (22) the Fr\'echet derivative of $K[u]$ can be obtained as 
\begin{equation}
D_P=\partial_{x}^{3}+\frac{3A}{5}(u\partial_{x}+u_x)\,\,.      
\end{equation}
 We shall, therefore, write 
\begin{equation}
M_3=a\partial_{x}^{3}+\frac{3A}{5}(bu\partial_{x}+cu_x)\,\,. 
\end{equation} 
Here the subscript $3$ on $M$ indicates that (30) represents the second Lax operator for the third-order equation. We shall follow this convention throughout. Equations (22), (27), (28) and (30) can be combined to get $a=4$, $b=1$ and $c=\frac{1}{2}$. Thus we have
\begin{equation}
\!\!\!\!\!\!\!\!\!\!\!\!\!\!\!\!\!\!\!\!\!\!\!\!\!\!\!\!\!\!\!\!\!\!\!\!\!\!\!\!\!\!\!\!\!\!\!\!\!\!\!\!\!\!\!\!\!\!\!\!\!\!\!\!\!\!\!\!\!\!\!\!\!\!\!\!\!\!\!\!\!\!\!\!\!\!\!\!\!\!\!\!\!\!\!\!\!\!\!\!\!\!\!\!\!\!\!\!\!\!\!\!M_3=4\partial_{x}^{3}+\frac{3A}{5}(u\partial_{x}+\frac{1}{2}u_x)\,\,.
\end{equation} 
Similarly, we find the results
\begin{eqnarray}
M_5=16\partial_{x}^{5}+4Au\partial_{x}^{3}+6Au_x\partial_{x}^{2}+5Au_{2x}\partial_x + \frac{3A^2}{10}u^2\partial_x+\frac{3A}{2}u_{3x}+\frac{3A^2}{10}uu_x\,\,,
\end{eqnarray}
\begin{eqnarray}
M_7=64\partial_{x}^{7}+\frac{112A}{5}u\partial_{x}^{5}+56Au_x\partial_{x}^{4}+84Au_{2x}\partial_{x}^{3}+\frac{14A^2}{5}u^2\partial_{x}^{3}+70Au_{3x}\partial_{x}^{2}\nonumber\\+\frac{42A^2}{5}uu_x\partial_{x}^{2}+\frac{161A}{5}u_{4x}\partial_x+7A^2uu_{2x}\partial_x+\frac{147A^2}{30}u_{x}^{2}\partial_x+\frac{7A^3}{50}u^3\partial_x\nonumber\\+\frac{63A}{10}u_{5x}+\frac{21A^2}{10}uu_{3x}+\frac{21A^2}{5}u_xu_{2x}+\frac{21A^3}{100}u^2u_x\,\,,
\end{eqnarray} 
and
\begin{eqnarray}
M_9=256\partial_{x}^{9}+\frac{576A}{5}u\partial_{x}^{7}+\frac{51A}{2}u_{7x}+\frac{2016A}{5}u_x\partial_{x}^{6}+\frac{897A}{5}u_{6x}\partial_{x}+\frac{4368A}{5}u_{2x}\partial_{x}^{5}\nonumber\\+\frac{2814A}{5}u_{5x}\partial_{x}^{2}+1176Au_{3x}\partial_{x}^{4}+\frac{5124A}{5}u_{4x}\partial_{x}^{3}+252A^2u_xu_{2x}\partial_{x}^{2}+\frac{5061A^2}{50}u_{2x}^{2}\partial_{x}\nonumber\\+\frac{546A^2}{5}u_{x}^{2}\partial_{x}^{3}+\frac{3654A^2}{25}u_xu_{3x}\partial_{x}+\frac{756A^2}{5}uu_{2x}\partial_{x}^{3}+126A^2uu_{3x}\partial_{x}^{2}\nonumber\\+\frac{609A^2}{10}u_{2x}u_{3x}+\frac{504A^2}{25}u^2\partial_{x}^{5}+\frac{567A^2}{10}uu_{5x}+\frac{516A^2}{5}uu_{x}\partial_{x}^{4}+\frac{2967A^2}{50}uu_{4x}\partial_{x}\nonumber\\+\frac{1743A^2}{50}u_xu_{4x}+\frac{189A^3}{100}u_{x}^{3}+\frac{441A^3}{50}uu_{x}^{2}\partial_{x}+\frac{42A^3}{25}u^3\partial_{x}^{3}+\frac{21A^3}{100}u^2u_{3x}\nonumber\\+\frac{378A^3}{50}u^2u_x\partial_{x}^{2}+\frac{63A^3}{10}u^2u_{2x}\partial_{x}+\frac{189A^3}{25}uu_xu_{2x}+\frac{63A^4}{1000}u^4\partial_{x}+\frac{63A^4}{500}u^3u_x\,\,.
\end{eqnarray} 
\par Zakharov and Faddeev {[}20{]} developed the Hamiltonian approach to
integrability of nonlinear evolution equations in one spatial and one temporal
(1+1) dimension and, in particular, Gardner {[}21{]} interpreted the KdV
equation as a completely integrable Hamiltonian system with $\partial_x$ as
the relevant Hamiltonian operator. A significant development in the
Hamiltonian theory is due to Magri {[}11{]} who realized that integrable
Hamiltonian systems have an additional structure. They are bi-Hamiltonian i.e.
they are Hamiltonian with respect to two different compatible Hamiltonian
operators. The bi-Hamiltonian structure of the integrable equation is based on a mathematical formulation that does not make explicit reference to the
Lagrangian of the equations in the hierarchy {[}22{]}. Here we shall demonstrate that the bi-Hamiltonian structure of the system of equations (20)-(24) can be realized in terms of a set of Hamiltonian densities obtained from the Lagrangians. Using (4) we can obtain the Lagrangian densities for our equations. In particular, we have 
\begin{equation}
\!\!\!\!\!\!\!\!\!\!\!\!\!\!\!\!\!\!\!\!\!\!\!\!\!\!\!\!\!\!\!\!\!\!\!\!\!\!\!\!\!\!\!\!\!\!\!\!\!\!\!\!\!\!\!\!\!\!\!\!\!\!\!\!\!\!\!\!\!\!\!\!\!\!\!\!\!\!\!\!\!\!\!\!\!\!\!\!\!\!\!\!\!\!\!\!\!\!\!\!\!\!\!\!\!\!\!\!\!\!\!\!\!\!\!\!\!\!\!\!\!{\cal L}_1=\frac{1}{2}w_tw_x-\frac{1}{2}w_{x}^{2}\,\,,
\end{equation}
\begin{equation}
\!\!\!\!\!\!\!\!\!\!\!\!\!\!\!\!\!\!\!\!\!\!\!\!\!\!\!\!\!\!\!\!\!\!\!\!\!\!\!\!\!\!\!\!\!\!\!\!\!\!\!\!\!\!\!\!\!\!\!\!\!\!\!\!\!\!\!\!\!\!\!\!\!\!\!\!\!\!\!\!\!\!\!\!\!\!\!\!\!\!\!\!{\cal L}_3=\frac{1}{2}w_tw_x-\frac{1}{2}w_xw_{3x}+\frac{A}{10}w_{x}^{3}\,\,,
\end{equation} 
\begin{equation}
\!\!\!\!\!\!\!\!\!\!\!\!\!\!\!\!\!\!\!\!\!\!\!\!\!\!\!\!\!\!\!\!\!\!{\cal L}_5=\frac{1}{2}w_tw_x-\frac{1}{2}w_xw_{5x}+\frac{A}{3}w_{x}^{2}w_{3x}+\frac{A}{6}w_xw_{2x}^{2}-\frac{A^2}{40}w_{x}^{4}\,\,,
\end{equation}
\begin{eqnarray}
\!\!\!\!\!\!\!{\cal L}_7=\frac{1}{2}w_tw_x-\frac{1}{2}w_xw_{7x}+\frac{7A}{10}w_xw_{3x}^{2}-\frac{7A^2}{40}w_{x}^{2}w_{2x}^{2}-\frac{7A^2}{40}w_{x}^{3}w_{3x}\nonumber\\\!\!\!\!\!\!\!\!\!\!\!\!\!\!\!\!\!\!\!\!\!\!\!\!\!\!\!\!\!\!\!\!\!\!\!\!\!\!\!\!\!\!\!\!\!\!\!\!\!\!\!\!\!\!\!\!\!\!\!\!\!\!\!\!\!\!\!\!\!\!\!\!\!\!\!\!\!\!\!\!\!\!\!\!\!\!\!\!\!\!\!\!\!\!\!\!\!\!\!\!\!\!\!\!\!\!\!\!\!\!\!\!\!\!\!\!\!\!\!\!\!+\frac{7A^3}{1000}w_{x}^5\,\,,
\end{eqnarray}
and
\begin{eqnarray}
\,\,\,\,{\cal L}_9=\frac{1}{2}w_tw_x-\frac{1}{2}w_xw_{9x}-\frac{3A}{5}w_{2x}^{2}w_{5x}+\frac{8A}{5}w_{3x}^{3}-\frac{9A}{10}w_xw_{4x}^{2}+\frac{7A^2}{40}w_{2x}^{4}\nonumber\\\!\!\!\!+\frac{63A^2}{200}w_{x}^{2}w_{3x}^{2}-\frac{21A^3}{100}w_{x}^{3}w_{2x}^{2}-\frac{21A^4}{10000}w_{x}^6.
\end{eqnarray}
In the above ${\cal L}_1$ is the Lagrangian density for the linear equation in (21). The other subscripts on $\cal L$ are self explanatory. The corresponding Hamiltonian densities are given by
\begin{equation}
\!\!\!\!\!\!\!\!\!\!\!\!\!\!\!\!\!\!\!\!\!\!\!\!\!\!\!\!\!\!\!\!\!\!\!\!\!\!\!\!\!\!\!\!\!\!\!\!\!\!\!\!\!\!\!\!\!\!\!\!\!\!\!\!\!\!\!\!\!\!\!\!\!\!\!\!\!\!\!\!\!\!\!\!\!\!\!\!\!\!\!\!\!\!\!\!\!\!\!\!\!\!\!\!\!\!\!\!\!\!\!\!\!\!\!\!\!\!\!\!\!\!\!\!\!\!\!\!\!\!\!\!\!\!\!\!\!\!\!\!\!\!\!\!\!\!{\cal H}_1=\frac{1}{2}u^2\,\,,
\end{equation}
\begin{equation}
\!\!\!\!\!\!\!\!\!\!\!\!\!\!\!\!\!\!\!\!\!\!\!\!\!\!\!\!\!\!\!\!\!\!\!\!\!\!\!\!\!\!\!\!\!\!\!\!\!\!\!\!\!\!\!\!\!\!\!\!\!\!\!\!\!\!\!\!\!\!\!\!\!\!\!\!\!\!\!\!\!\!\!\!\!\!\!\!\!\!\!\!\!\!\!\!\!\!\!\!\!\!\!\!\!\!\!\!\!\!\!\!\!\!\!\!\!\!\!{\cal H}_3=\frac{1}{2}uu_{2x}+\frac{A}{10}u^3\,\,,
\end{equation}
\begin{equation}
\!\!\!\!\!\!\!\!\!\!\!\!\!\!\!\!\!\!\!\!\!\!\!\!\!\!\!\!\!\!\!\!\!\!\!\!\!\!\!\!\!\!\!\!\!\!\!\!\!\!\!\!\!\!\!\!\!\!\!\!\!\!\!\!\!\!\!\!\!\!\!{\cal H}_5=\frac{1}{2}uu_{4x}+\frac{A}{3}u^2u_{2x}+\frac{A}{6}uu_{x}^{2}+\frac{A^2}{40}u^4\,\,,
\end{equation}
\begin{equation}
\!\!\!\!\!\!\!\!\!\!\!\!\!\!\!\!\!\!\!\!\!\!\!\!\!\!{\cal H}_7=\frac{1}{2}uu_{6x}+\frac{7A}{10}uu_{2x}^{2}+\frac{7A^2}{40}u^2u_{x}^{2}+\frac{7A^2}{40}u^3u_{2x}+\frac{7A^3}{1000}u^5\,\,,
\end{equation}
and 
\begin{eqnarray}
\,\,\,\,\,\,\,{\cal H}_9=\frac{1}{2}uu_{8x}-\frac{3A}{5}u_{x}^{2}u_{4x}+\frac{8A}{5}u_{2x}^{3}-\frac{9A}{10}uu_{3x}^{2}-\frac{63A^2}{200}u^2u_{2x}^{2}-\frac{7A^2}{40}u_{x}^{4}\nonumber\\-\frac{21A^3}{100}u^3u_{x}^{2}+\frac{21A^4}{10000}u^6.
\end{eqnarray}
\par In the theory of Zakharov and Faddeev {[}20{]} and of Gardner {[}21{]} the Hamiltonian form of an integrable nonlinear evolution equation reads
\begin{equation}
\!\!\!\!\!\!\!\!\!\!\!\!\!\!\!\!\!\!\!\!\!\!\!\!\!\!\!\!\!\!\!\!\!\!\!\!\!\!\!\!\!\!\!\!\!\!\!\!\!\!\!\!\!\!\!\!\!\!\!\!u_t=\partial_x(\frac{\delta\cal H}{\delta u})
\end{equation}
with $\cal H$, the Hamiltonian densities of that equation. Here $\frac{\delta}{\delta u}$ denotes the usual variational derivative written as
\begin{equation}
\frac{\delta}{\delta u}=\sum_{n\geq 0}(-\partial_x)^n\frac{\partial}{\partial
u_n} ,\,\,\,\,u_n=(\partial_x)^nu\,\,.
\end{equation}
Using the Hamiltonian densities in (40)-(44), one can easily verify the
Faddeev-Zakharov-Gardner equation in (45) to yield the appropriate nonlinear equations in (20)-(24). The bi-Hamiltonian form of evolution equations is given by {[}11{]}
\begin{equation}
\!\!\!\!\!\!\!\!\!\!\!\!\!u_t=\partial_x(\frac{\delta {\cal H}_{m+2}}{\delta u})={\cal E}(\frac{\delta {\cal H}_{m}}{\delta u})
\end{equation}
with $m=2n+1,\,\,\, n=0,1,2,.......\,\,\,\,.$ In (47) the second Hamiltonian
operator is related to the recursion operator by {[}16{]}
\begin{equation}
\!\!\!\!\!\!\!\!\!\!\!\!\!\!\!\!\!\!\!\!\!\!\!\!\!\!\!\!\!\!\!\!\!\!\!\!\!\!\!\!\!\!\!\!\!\!\!\!\!\!\!\!\!\!\!\!{\cal E}={\Lambda }{\partial_x}
\end{equation} 
From (15) and (48) we get
\begin{equation}
\!\!\!\!\!\!\!\!\!\!\!\!\!\!\!{\cal E}=
\partial_{x}^{3}+\frac{2A}{5}u\partial_x+\frac{A}{5}u_x\,\,.
\end{equation}
From $(47)$ and $(49)$ we have
\begin{equation}
u_t=\partial_x(\frac{\delta {\cal H}_{m+2}}{\delta
u})=(\partial_{x}^{3}+\frac{2A}{5}u\partial_x+\frac{A}{5}u_x)(\frac{\delta{\cal H}_{m}}{\delta u})\,\,.
\end{equation}
For $n=1$ (50) reads
\begin{equation}
\!\!\!\!\!\!\!\!\!\!\!u_t=\partial_x(\frac{\delta {\cal H}_{5}}{\delta
u})=(\partial_{x}^{3}+\frac{2A}{5}u\partial_x+\frac{A}{5}u_x)(\frac{\delta{\cal H}_{3}}{\delta u})\,\,.
\end{equation}
From (41), (42) and (51) one can easily obtain (20) verifying the bi-Hamiltonian structure. Similar results can also be checked for other pairs of the Hamiltonians in (40)-(44).
\vskip 0.2 cm
\noindent {\bf\large IV. Soliton solution}
 \vskip 0.3 cm
 We have just seen that the bi-Hamiltonian form $(51)$ corresponds to the fifth-order nonlinear equation in $(20)$. Here we shall make use of the homogeneous
 balance method (HB) {[}15{]} to construct an analytical expression for the soliton
 solution of this equation. According to HB method, the field variable is
 first expanded as
 $$\!\!\!\!\!\!\!\!\!\!\!\!\!\!\!\!\!\!\!\!\!\!\!\!\!\!\!\!\!\!\!\!\!\!\!\!\!\!\!\!\!\!\!\!\!\!\!\!\!\!\!\!\!\!\!\!u(x,t)=\sum_{i=0}^{N}f^{(i)}\left(w(x,t)\right)\,\,,\eqno(52)$$
 where the superscript $(i)$ denotes the derivative index. In particular,
 $f^{(1)}=\frac{\partial f}{\partial w},\,\,f^{(2)}=\frac{\partial^2 f}{\partial
 w^2} $ and so on. Substituting $(52)$ in $(20)$ and balancing the
 contribution of the linear term with that of the nonlinear terms, the
 expression in $(52)$ becomes restricted to
 $$\!\!\!\!\!\!\!\!\!\!\!\!\!\!\!\!\!\!\!\!\!\!\!\!\!\!\!\!\!\!\!\!\!\!\!\!\!\!\!\!\!\!\!\!\!\!\!\!\!\!\!\!\!\!\!\!u(x,t)=f^{(2)}w_{x}^{2}+f^{(1)}w_{2x}\,\,,\eqno(53)$$
 where subscripts on $w$ stand for appropriate partial derivative. From $(53)$
 and $(20)$ we have 
 $$(f^{(7)}+Af^{(2)}f^{(5)}+2Af^{(3)}f^{(4)}+\frac{3A^2}{10}(f^{(2)})^2f^{(3)})w_{x}^{7}\eqno(54)$$
$+${\it other terms involving lower powers of the partial derivatives of} $w=0\,\,.$
Setting the coefficient of $w_{x}^{7}$ to zero we get
$$f^{(7)}+Af^{(2)}f^{(5)}+2Af^{(3)}f^{(4)}+\frac{3A^2}{10}(f
^{(2)})^2f^{(3)}=0\,\,.\eqno(55)$$
If we try a solution of $(55)$ in the form
$$f=\alpha \,\,lnw\eqno(56)$$
we immediately get
$$\alpha={20\over A}\,\,.\eqno(57)$$
From $(56)$ we can deduce the following results
$$\!\!\!\!\!\!\!\!\!\!\!\!\!\!\!\!\!\!\!\!\!\!\!\!\!f^{(2)}f^{(5)}=-{\alpha\over30}f^{(7)},\,\,f^{(3)}f^{(4)}=-{\alpha\over60}f^{(7)},\,\,f^{(2)})^2f^{(3)}={{\alpha^2}\over360}f^{(7)};$$$$\!\!\!\!\!\!\!\!\!\!\!\!\!\!\!\!\!\!\!\!\!\!\!\!\!\!\!f^{(2)}f^{(4)}=-{\alpha\over20}f^{(6)},\,\,(f^{(3)})^2=-{\alpha\over30}f^{(6)},\,\,f^{(1)}f^{(5)}=-{\alpha\over5}f^{(6)},\,\,$$$$(f^{(2)})^3={{\alpha^2}\over120}f^{(6)},\,\,f^{(1)}f^{(2)}f^{(3)}={{\alpha^2}\over60}f^{(6)};$$$$\!\!\!\!\!\!\!\!\!\!\!\!\!\!\!\!\!\!\!f^{(2)}f^{(3)}=-{\alpha\over12}f^{(5)},\,\,f^{(1)}f^{(4)}=-{\alpha\over4}f^{(5)},\,\,\left(f^{(2)}\right)^2f^{(1)}={{\alpha^2}\over24}f^{(5)},\,\,$$$$(f^{(1)})^2f^{(3)}={{\alpha^2}\over12}f^{(5)};$$$$\!\!\!\!\!\!\!\!\!\!\!\!\!\!\!\!\!\!\!\!\!\!\!\!\!\!\!\!f^{(1)}f^{(3)}=-{\alpha\over3}f^{(4)},\,\,(f^{(2)})^2=-{\alpha\over6}f^{(4)},\,\,(f^{(1)})^2f^{(2)}={{\alpha^2}\over6}f^{(4)};$$$$\!\!\!\!\!\!\!\!\!\!\!\!\!\!\!\!
\!\!\!\!\!\!\!\!\!\!\!\!\!\!\!\!\!\!\!\!\!\!\!\!\!\!\!\!\!\!\!\!\!\
\!\!\!\!\!\!\!\!\!\!\!\!\!\!\!\!\!\!\!\!\!\!\!\!\!\!\!\!\!\!\!\!\!\!\!\!
\!\!f^{(1)}f^{(2)}=-{\alpha\over2}f^{(3)},\,\,(f^{(1)})^3={{\alpha^2}\over2}f^{(3)};$$$$\!\!\!\!\!\!\!\!\!\!\!\!\!\!\!\!\!\!\!\!\!\!\!\!\!\!\!\!\!\!\!\!\!\!\!\!\!\!\!\!\!\!\!\!\!\!\!\!\!\!\!\!\!\!\!\!\!\!\!\!\!\!\!\!\!\!\!\!\!\!\!\!\!\!\!\!\!\!\!\!\!\!\!\!\!\!\!\!\!\!\!\!\!\!\!\!\!\!\!\!\!\!\!\!\!\!\!\!\!\!\!\!\!\!\!\!\!\!\!\!\!\!\!\!\!\!\!\!\!\!\!(f^{(1)})^2=-\alpha
f^{(2)}\,\,.\eqno(58)$$
Substituting $(58)$ in the full form of $(54)$, the latter is reduced to a
linear polynomial in $f^{(1)},\,\,f^{(2)},....,f^{(7)}\,\,.$ If the
coefficient of each $f^{(i)}$ is set equal to zero we get a set of partial
differential equations for $w(x,t)$
$$\!\!\!\!\!\!\!\!\!\!\!\!\!\!\!\!\!\!\!\!\!\!\!\!\!\!\!\!\!\!\!\!\!\!\!\!\!\!\!\!\!\!\!\!\!\!\!\!\!\!\!\!\!\!\!\!\!\!\!\!\!\!\!\!\!\!\!\!\!\!\!\!\!\!\!\!\!\!\!\!\!\!\!\!\!\!\!\!\!\!\!\!\!\!\!\!\!\!\!\!\!\!\!\!\!\!\!\!\!\!\!\!\!\!\!\!\!\!\!\!\!\!\!\!\!\!\!\!\!\!\!\!\!\!\!\!\!\!\!\!\!\!\!\!\!\!\!w_{xxt}-w_{7x}=0\,\,,\eqno(59a)$$
$$2w_xw_{xt}+w_tw_{xt}+(2A\alpha-35)w_{3x}w_{4x}+(A\alpha-21)w_{2x}w_{5x}-7w_xw_{6x}=0\,\,,\eqno(59b)$$
$$\!\!\!\!\!\!\!2w_tw_{x}^{2}+(A\alpha-42)w_{x}^{2}w_{5x}+(11A\alpha-210)w_xw_{2x}w_{4x}+(8A\alpha-140)w_xw_{3x}^{2}$$$$\!\!\!\!\!\!\!\!\!\!\!\!\!\!\!\!\!\!\!\!\!\!\!\!\!\!\!\!\!\!\!\!\!\!\!\!\!\!\!\!\!\!\!\!\!\!\!\!\!\!\!+(16A\alpha-\frac{3A^2}{10}\alpha^2-210)w_{2x}^{2}w_{3x}=0\,\,,\eqno(59c)$$
$$\!\!\!\!\!\!\!\!\!\!\!\!\!\!\!\!\!\!\!\!\!\!\!(48A\alpha-\frac{9A^2}{10}\alpha^2-630)w_{x}w_{2x}^{3}+(78A\alpha-\frac{3A^2}{5}\alpha^2-1260)w_{x}^{2}w_{2x}w_{3x}$$$$\!\!\!\!\!\!\!\!\!\!\!\!\!\!\!\!\!\!\!\!\!\!\!\!\!\!\!\!\!\!\!\!\!\!\!\!\!\!\!\!\!\!\!\!\!\!\!\!\!\!\!\!\!\!\!\!\!\!\!\!\!\!\!\!\!\!\!\!+(9A\alpha-210)w_{x}^{3}w_{4x}=0\,\,,\eqno(59d)$$
$$(174A\alpha-\frac{12A^2}{5}\alpha^2-2520)w_{x}^{3}w_{2x}^{2}+(48A\alpha-\frac{3A^2}{10}\alpha^2-840)w_{x}^{4}w_{3x}=0\,\,,\eqno(59e)$$
$$\!\!\!\!\!\!\!\!\!\!\!\!\!\!\!\!\!\!\!\!\!\!\!\!\!\!\!\!\!\!\!\!\!\!\!\!\!\!\!\!\!\!\!\!\!\!\!\!\!\!\!\!\!\!\!\!\!\!\!\!\!\!\!\!\!\!\!\!\!\!\!\!\!\!\!\!\!\!\!\!\!\!\!\!\!\!\!\!\!\!\!\!\!\!\!\!\!\!\!\!\!\!\!\!\!(24A\alpha-\frac{3A^2}{10}\alpha^2-360)w_{x}^{5}w_{2x}=0\eqno(59f)$$
and
$$\!\!\!\!\!\!\!\!\!\!\!\!\!\!\!\!\!\!\!\!\!\!\!\!\!\!(24A\alpha-\frac{3A^2}{10}\alpha^2-360)w_{x}^{7}=0\,\,.\eqno(59g)$$
\par Equation $(59a)$ is a linear partial differential equation and can be
converted to an ordinary differential equation by substituting
$$\!\!\!\!\!\!\!\!\!\!\!\!\!\!\!w(x,t)=g(x+vt)=g(z)\,\,.\eqno(60)$$
Using $(60)$ in $(59a)$ we have
$$\!\!\!\!\!\!\!\!\!\!\!\!\!\!\!\!\!\!\!\!\!\!\!\!\!\!\!\!\!\!\!\!\!\!\!\!\!\!v\frac{d^3g}{dz^3}-\frac{d^7g}{dz^7}=0\,\,.\eqno(61)$$
Here $v$ is the velocity of the travelling wave represented by $w(x,t)$.
Equation $(61)$ can be solved to write 
$$\,\,\,\,\,\,\,\,\,\,w(x,t)=g(x+vt)=c_0+c_1e^{{\sqrt[4]v}(x+vt)}\,\,,\eqno(62)$$
where $c_0$ and $c_1$ are arbitrary constants. Using $(56)$, $(57)$ and $(62)$
in $(53)$ we get the exact soliton solution of the fifth-order equation in
$(20)$ and/or $(51)$ in the form 
$$u_5(x,t)={20\over A}\frac{c_0c_1{\sqrt v}e^{{\sqrt[4]v}(x+vt)}}{(c_0+c_1e^{{\sqrt[4]v}(x+vt)})^2}\,\,.\eqno(63)$$
A similar result for the third-order equation in $(22)$ is given by
$$u_3(x,t)={20\over A}\frac{c_0c_1ve^{{\sqrt v}(x+vt)}}{(c_0+c_1e^{{\sqrt v}(x+vt)})^2}\,\,.\eqno(64)$$
The subscripts on $u(x,t)$ are self explanatory. It is of interest to note
that for $c_1=c_0=1$, $A=10$ and $v=4\kappa^2$, $u_3(x.t)$ in $(64)$ becomes
$$u_3(x,t)=2\kappa^2sech^2(\kappa x+4\kappa^3t)\,\,.\eqno(65)$$
From the inverse spectral method {[}23{]} for solving the KdV equation, we
know that $\kappa^2$ has a simple physical meaning. For example $-\kappa^2$
represents a discrete energy eigenvalue of the Sch\"odinger equation for the
initial potential $u_3(x,0)$. As in ref. 9 we shall now examine the spatial
behaviour of $u_5(x,t)$ at $t=0$. For the sake of simplicity we shall work
with $v=1$. In Fig.1 we plot $u_5(x,0)$ as function of $x$ for different
values of parameter $c_0$ and $c_1$.
\begin{figure}
{\centering\resizebox*{0.8\textwidth}{!}{\rotatebox{-90}{\includegraphics{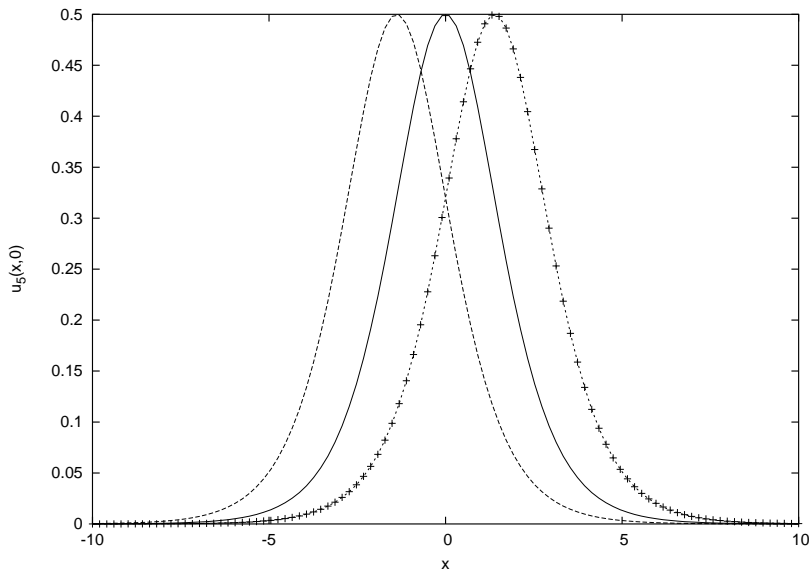}}}\par}
\caption{Variation of $u_5(x,0)$ with $x$}
\end{figure}
 All the curves in the figure are of
$sech^2$ shape indicating that solution obtained from $(63)$ have indeed
solitary wave properties. The solid curve for $c_0=1$ and $c_1=1$ is centred
at the point $x=0$. If $c_0$ and $c_1$ are made unequal, the centre of the
soliton moves either to the left or to the right. In particular, for
$c_0>c_1$, the shift of the centre is towards the right and we have a reverse
situation for $c_0<c_1$. We have displayed this property by using a dashed
curve with cross
$(c_0=4$ and $c_1=1)$ and a simple dashed curve $(c_0=1$ and $c_1=4)$.
 \vskip 0.2 cm
 \noindent {\bf\large V. Conclusion}
 \vskip 0.3 cm
 \par Fifth-order nonlinear evolution equations, on the one hand, have a host
 of connections with other important integrable equations and, on the other
 hand, can not be solved by simple analytical methods. These two points
 inspired us to construct a general fifth-order equation which follow from a
 Lagrangian. It is often desirable that equations of mathematical physics
 should be derivable from an action principle because a non-Lagrangian system
 does not allow one to carry out a linear stability check {[}24{]} as well as
 to derive a field theory {[}25{]} for particles described by its solutions.
 The Lagrangian approach to nonlinear evolution equations is quite interesting
 because here one can derive all physico-mathematical results from first
 principles {[}8{]}. Based on the fifth-order Lagrangian equation we derived
 an integrable hierarchy. As a test of integrability we provided a Lax
 representation and constructed two compatible Hamiltoinan structures.\par We
 treated the third- and fifth-order equations in the hierarchy by the
 homogeneous balance method {[}15{]} to obtain analytical results for soliton
 solutions. Ideally, we could have tried the bi-linear method of Hirota
 {[}26{]} to deal with the problem because this method is very convenient for
 finding single- and multi-soliton solutions of nonlinear evolution equations.
 For higher-order equations, Hirota transformation often leads to multilinear
 representation {[}27{]}. This tends to pose problems in solving the equations. The homogeneous balance method, on the other hand, does not involve any new mathematical complication as one moves from lower- to higher-order equations. Admittedly, the algebra become more and more involved as we go up the ladder inside the hierarchy. The symbolic computations facilities like the Mapple and Mathematica can be used to circumvent algebraic complications. 
\vskip 1.2 cm
Acknowledgement. This work is supported by the University Grants Commission, Government of India, through grant No. F.10-10/2003(SR). 
\newpage
{[}1{]} K. Nakkeeran, K. Porsezian, P. Shanmugha Sundaram and A. Mahalingam,  Phys. Rev. Lett.  {\bf 80}, 1425 (1998).\\

{[}2{]} U. Al Khawaja, H. T. C. Stoof, R. G. Hulet, K. E. Strecker, and G. B. Patridge,  Phys. Rev. Lett.  {\bf 89}, 200404 (2002).\\

{[}3{]} M. N. \"Ozer and F. T. D\"oken, J. Phys. A.: Math. Gen.  {\bf 36}, 2319 (2003).\\

{[}4{]} M. N. \"Ozer and I. Da\u{g}, Hadronic. J.  {\bf 24}, 195 (2001).\\

{[}5{]} P. D. Lax, Comm.  Pure  Appl. Math.  {\bf 21}, 467 (1986).\\

{[}6{]} K. Sawada and T. Kotera,  Prog. Theo. Phys.  {\bf 51}, 1355 (1974).\\

{[}7{]} D. J. Kaup, Stud.Appl.Math.  {\bf 62}, 189 (1980); B. Kupersmidt, Commun. Math. Phys. {\bf 99}, 51 (1988).\\

{[}8{]} P. Rosenau and J. M. Hymann, Phys. Rev. Lett.  {\bf 70}, 564 (1993);
B. Talukdar, J. Shamanna and S. Ghosh, Pramana. J. Phys.  {\bf 61}, 99 (2003);
S. Ghosh, U. Das and B. Talukdar, Int. J. Theor. Phys.  {\bf 44}, 363 (2005)\\

{[}9{]} W. Hong and Y. Jang, Z. Naturforsch.  {\bf 54a}, 549 (1999).\\

{[}10{]}  R. M. Santili, Foundations of Theoretical Mechanics I (Springer-Verlag, New York, 1984).\\

{[}11{]} F. Magri, J. Math. Phys.  {\bf 19}, 1548 (1971).\\

{[}12{]} B. Talukdar, S. Ghosh, J. Shamanna and P. Sarkar, Eur. Phys. J. {\bf D21}, 105 (2002); B. Talukdar. S. Ghosh and U. Das, J. Math. Phys.  {\bf 46}, 043506 (2005).\\

{[}13{]} D. Anderson, Phys. Rev. A.  {\bf 27}, 3135 (1983).\\

{[}14{]} F. Cooper, C. Lucheroni, H. Shepard and P. Sodano, arxiv:hep-ph/9210226 v1, 9 Oct 92. \\ 

{[}15{]} M. Wang, Phys. Lett. A. {\bf 199}, 169 (1995); ibid {\bf 213}, 279 (1996); {\bf 216}, 67 (1996).\\

{[}16{]} P. J. Olver, Application of Lie Groups to Differential Equation, (Springer-Verlag, NY, 1993).\\

{[}17{]} M. Ito, J. Phys. Soc. Japan  {\bf 49}, 771 (1980).\\

{[}18{]} F. Calogero and A. Degasperis, Spectral Transform and Soliton (North-Holland Publising Company, New York, 1982).\\

{[}19{]} K. Chadan and P. C. Sabatier, Inverse problems in Quantum Scattering Theory (2nd ed, Springer, New York, 1989).\\

{[}20{]} V. E. Zakharov and L. D. Faddeev, Funct. Anal. Phys. {\bf 5}, 18 (1971).\\

{[}21{]} C. S. Gardner, J. Math. Phys. {\bf 12}, 1548 (1971).\\

{[}22{]} S. Ghosh, B. Talukdar and J. Shamanna, Czech. J.  Phys. {\bf 53}, 425 (2003).\\

{[}23{]} C. S. Gardner, J. M. Greene, M. D. Kruskal and R. M. Miura, Phys. Rev. Lett. {\bf 19}, 1095 (1967).\\

{[}24{]} B. Dey and A. Khare, J. Phys. {\bf 33}, 5335 (2000).\\

{[}25{]} S. A. Hojman and L. C. Shepley, J. Math. Phys. {\bf 32}, 142 (1990).\\ 

{[}26{]} R. Hirota, Phys. Rev. Lett. {\bf 27}, 1192 (1971).\\

{[}27{]} J. Hietarinta in Nonlinear Dynamics: Integrability and Chaos ed. M. Daniel, K. M. Tamizhmani and R. Sahadevan (Narosa Publising House, New Delhi (2000)).
\end{document}